\documentclass[12pt]{article}

\usepackage[a4paper, margin=1in]{geometry}

\usepackage{mathptmx}

\usepackage[compact]{titlesec}
\titlespacing{\section}{0pt}{2ex}{1ex}
\titlespacing{\subsection}{0pt}{1ex}{0ex}
\titlespacing{\subsubsection}{0pt}{0.5ex}{0ex}

\usepackage[utf8]{inputenc} 
\usepackage[T1]{fontenc}    
\usepackage{hyperref}       
\usepackage{url}            
\usepackage{nicefrac}       
\usepackage{graphicx}

\usepackage{amsmath,amssymb,amsfonts,amsthm}

\makeatletter
\g@addto@macro\normalsize{%
	\setlength\abovedisplayskip{5pt}
	\setlength\belowdisplayskip{5pt}
	\setlength\abovedisplayshortskip{5pt}
	\setlength\belowdisplayshortskip{5pt}
}

\usepackage{wrapfig}

\usepackage{caption}
\usepackage{accents}

\usepackage{subcaption}

\usepackage{setspace}
 \usepackage[backend=bibtex8,url=true,eprint=true,doi=true,sorting=none,backref=true]{biblatex}
\AtEveryBibitem{
	\clearfield{urlyear}
	\clearfield{urlmonth}
}

\newcommand{\bite}{\begin{itemize}}
	\newcommand{\eat}{\end{itemize}}
\newcommand{\beq}{\begin{equation}}
	\newcommand{\eeq}{\end{equation}}

\newcommand{\beqa}{\begin{align}}
	\newcommand{\eeqa}{\end{align}}
\newcommand{\barr}{\begin{array}}
	\newcommand{\earr}{\end{array}}

\newcommand{\C}{\mathbb{C}}

\newcommand{\bz}{\mathbf{z}}

\newcommand{\mb}[1]{\mathbf{#1}}
\newcommand{\mc}[1]{\mathcal{#1}}
\newcommand{\mbb}[1]{\mathbb{#1}}
\newcommand{\mf}[1]{\mathfrak{#1}}

\usepackage{bbold}

\newcommand{\id}{\mathbb{1}}

\newcommand{\vect}[1]{\boldsymbol{#1}}
\newcommand{\bvec}[1]{\boldsymbol{\vec #1}}
\newcommand{\expect}[1]{\langle #1\rangle}
\newcommand{\innerp}[2]{\langle #1 \vert #2 \rangle}

\newcommand{\bra}[1]{\langle #1 \vert}
\newcommand{\ket}[1]{\vert #1 \rangle}

\newcommand{\sltwoc}{\mathfrak{sl}(2,\mathbb{C})}

\newcommand{\bket}[1]{\vert #1 )}

\title{Quantum Error Correction and Emergence of Stable Spacetime}

\author{
  Deepak Vaid \\
  Department of Physics, NITK Surathkal\\
  Mangaluru, Karnataka - 575025\\
  India\\
  \texttt{dvaid79@gmail.com} \\
}

\begin{document}

\begin{titlepage}
	\begin{center}
		\vspace*{1cm}
		
		\textbf{Lorentz Invariance, Scattering Amplitudes and the Emergence of Semiclassical Geometry}
		
		\vspace{0.5cm}
		Essay written for the Gravity Research Foundation 2022 Awards for Essays on Gravitation
		
		\vspace{1cm}
		
		\textbf{Deepak Vaid}
		
		\textbf{dvaid79@gmail.com}
		
		\begin{abstract}
			It has been known for some time now that error correction plays a fundamental role in the determining the emergence of semiclassical geometry in quantum gravity. In this work I connect several different lines of reasoning to argue that this should indeed be the case. The kinematic data which describes the scattering of $ N $ massless particles in flat spacetime can put in one-to-one correspondence with coherent states of quantum geometry. These states are labeled by points in the Grassmannian $ Gr_{2,n} $, which can be viewed as labeling the code-words of a quantum error correcting code. The condition of Lorentz invariance of the background geometry can then be understood as the requirement that co-ordinate transformations should leave the code subspace unchanged. In this essay I show that the language of subsystem (or operator) quantum error correcting codes provides the proper framework for understanding these aspects of particle scattering and quantum geometry.
		\end{abstract}
	
		Submission Date: Mar 30, 2022

		\vspace{0.8cm}
		
		Department of Physics\\
		Rm 217, Science Block\\
		NITK Surathkal\\
		Mangaluru\\
		Karnataka, India - 575025\\
		
	\end{center}
\end{titlepage}	
	


\section{Introduction}

It has been known for some time now that quantum error correcting codes should play a fundamental role in determining the emergence of smooth, semiclassical geometry from some underlying non-perturbative theory of quantum gravity. Arguments supporting this relationship center around the AdS/CFT correspondence. It is not clear, in particular, how one can extend the constructions of Pastawski, Almhieri, Harlow and collaborators \cite{Almheiri2014Bulk,Harlow2017The-RyuTakayanagi,Pastawski2015Holographic} (see also \cite{Vaid2019Quantum}) to asymptotically flat spacetimes such as Minkowski. Another issue is that these constructions, so far, consider the question of reconstruction of bulk geometries given the knowledge of conformal fields living on the boundary but do not address what, if any, role quantum error correction could (or should) play in describing matter degress of freedom and their interactions.

It seems natural to consider the proposition that, if geometry is indeed described by an error correcting code, then matter degrees of freedom and the interactions between matter and between matter and geometry should also have a description in terms of error correcting codes. In this essay I argue that this expectation is indeed borne out, at least when one considers the scattering of $ N $ massless particles in flat spacetime. There are three ingredients we require for this picture. First, an understanding of the structure of the kinematic space of massless $N$ particle scattering in terms of the Grassmannian. Second, the description of semiclassical states of geometry in terms of coherent intertwiners. And, finally, an understanding of the formalism of operators quantum error correcting codes (QECC). Before we dive into the details, however, I can state the end result so as to give the reader an idea of what to expect going forwards.

Equivalence classes of configurations of massless $ N $ particle scattering are related to each other by Lorentz transformations. Each equivalence class is labeled by a point in the complex Grassmannian $ Gr_{2,n} $. Lorentz transformations take us between different configurations within the same equivalence class. Hence the action of Lorentz transformations leaves points on the Grassmanian invariant. In the language of subsystem quantum error correction codes \cite{Kribs2005Unified} (also known as \emph{operator}  or \emph{gauge} QECC) Lorentz transformations can therefore be understood as gauge transformations which leave the code subspace invariant. Furthermore each individual equivalence class (consisting of configurations of kinematic data related to each other by Lorentz transformations), each of which corresponds to a single point on the Grassmannian, can then be viewed as codewords of this QECC.

\section{Subsystem Quantum Error Correction}

In this section I briefly review the idea behind subsystem or operator QECC (OQECC). In any error correcting code, the goal is to encode logical information in a code Hilbert space $ \mc{H}_C $ which is a subset of a bigger physical Hilbert space $ \mc{H}_C \in \mc{H}_P $. 

The general framework for subsystem or operator quantum error correcting codes is as follows. We can decompose $ \mc{H}_P $ into a direct sum:
\begin{equation}
	\mc H_P = C \oplus C^\perp
\end{equation}
where $ C $ is the code subspace and $ C^\perp $ is its orthogonal complement in $ \mc H_P $. Further the code subspace can be written as a tensor product:
\begin{equation}\label{key}
	C = A \otimes B
\end{equation}
where $ A $ is the logical Hilbert space in which our physical message is encoded and $ B $ encodes the gauge degrees of freedom of the code subspace. There exists a group of transformations $ \mc L $ which act only on $ B $ and not on $ A $ and leave the codewords invariant. These are known as ``gauge'' transformations for obvious reasons. Errors $ \mc E $ are a subset of the set of maps $ \mc B(\mc H_P) \rightarrow \mc B(\mc H_P)$, where $ \mc B (\mc H_P) $ is the set of bounded operators on $ \mc{H}_P $. We say that an error $ E_a \in \mc E $ is correctable if there exists a map $ \mc R: \mc B(\mc H_P) \rightarrow \mc B(\mc H_P) $ which reverses the action of $ E_a $ upto some transformation acting on the $ B $ subsystem. This condition can be expressed as follows \cite{Poulin2005Stabilizer}:
\begin{equation}\label{eqn:qec-condition}
	\mc R \circ \mc E (\rho_A \otimes \rho_B) = \rho_A \otimes \rho'_B
\end{equation}
where $ \rho_A \in \mc B(A) $ and $\rho_B, \rho'_B \in \mc B(B)$. If we express an arbitrary element of $ \mc E $ as $ \mc E = \sum_a E_a \rho E_a^\dagger $, then the above condition can be expressed as:
\begin{equation}\label{eqn:qec-condition-v2}
	PE^\dagger_a E_b P = \id^A \otimes g^B_{ab}
\end{equation}
where $ P $ is the projector onto the code-subspace: $ P \mc{H_P} = C $ and $ g^B_{ab} $ is an arbitrary quantum operation acting on $ B $. OQECC contains as a special case the so-called ``standard model'' of QECC (when $ \dim B = 1 $) and the framework of decoherence free subspaces \cite{Zanardi1997Noiseless,Kribs2005Geometry}.

\section{Scattering Amplitudes and the Grassmanian}

The scattering of elementary particles is the bread and butter of high energy physicists. It is the amplitudes associated with these processes which connect the esoteric techniques of quantum field theory (QFT) with the ``real'' world. The standard machinery for calculating scattering amplitudes involves the use of perturbation theory and the use of Feynman diagrams. Despite the tremendous effectiveness of these techniques\footnote{For instance, in the determination of the fine-structure constant $ \alpha = \frac{e^2}{hc} $ to a precision of eight significant digits.} the Feynman diagram method fails miserably when it comes to helping us understand the dynamics of particles and fields interacting via the strong force.

As an example one can consider interaction vertices for scattering between three of more gluons in the calculation of the amplitudes of most QCD processes. The evaluation of such amplitudes involves evaluating Feynman diagrams whose number increases exponentially with the total number of particles involved in a given process. However, beginning in the 1980s it began to be noticed that despite the apparently overwhelming complexity of these processes, after summing thousands of Feynman diagrams associated with a given process, the final result would collapse into a single expression of remarkable simplicity. This gave birth to the hope that there were some heretofore unknown symmetries which, if exploited correctly, could lead to dramatic simplifications in such calculations.

The canonical example of such a simplification is in the form of Parke-Taylor formula \cite{Parke1986Amplitude} \footnote{Introductions to the MHV formalism, BCFW recursion relations and related techniques can be found in several references. Some of these include \cite{Dixon1996Calculating,Alday2008Lectures,Feng2012An-Introduction,Elvang2013Scattering,Dixon2013A-brief}}:
\begin{equation}\label{eqn:parke-taylor}
	\mc{A}_n(1^+, \ldots, i^-, \ldots, j^-, \ldots, n^+) = \frac{\innerp{i}{j}^4}{\innerp{1}{2} \innerp{2}{3} \ldots \innerp{n-1}{n} \innerp{n}{1}}.
\end{equation}
for the tree-level amplitude for the scattering of $ n $ gluons. Here the numbers $ 1, \ldots, n $ represent the momenta of the $ n $ gluons written in terms of \emph{spinor helicity} variables and $ \innerp{i}{j} $ is the symplectic inner product of two spinors: $ \innerp{i}{j} = \epsilon^{ab}\lambda_{ia} \lambda_{jb} $, where $ a \in \{1,2\} $ labels the components of each spinor. The simplicity of this expression is striking especially when we consider that for tree-level processes involving $ n > 4 $ gluons, the number of Feynman diagrams which have to be summed over to obtain the final result grows exponentially as a function of $ n $.


\subsection{Spinor Helicity Formalism}\label{sec:spinor-helicity}

It is a well known fact that any vector $ p^\mu := (p^0,\vect{p}) $ in Minkowski spacetime can be mapped to Hermitian matrices as follows:
\begin{equation}\label{eqn:spinor-mapping}
	p_{\alpha \dot \alpha} = \sigma^\mu_{\alpha \dot \alpha} p_\mu
\end{equation}
where $ \sigma^\mu := (\id, \vect{\sigma}) $ are the three Pauli matrices supplemented with the identity matrix. It is easy to verify that the determinant of this matrix is equal to the squared norm of the associated 4-vector:
\begin{equation}\label{eqn:mom-determinant}
	\det{p} = p^\mu p_\mu = p^2
\end{equation}
Now, in order to describe the scattering of $ n $ particles we need to specify $ n $ momentum vectors: $ \{p_1^\mu, p_2^\mu, \ldots, p_n^\mu\} $. For massless particles the momenta must be null vectors: $ p^2 = 0 $. The corresponding matrices will therefore have vanishing determinant and will be of rank 1. Now any rank 1 matrix can be written as the product of two vectors as follows:
\begin{equation}\label{eqn:null-momenta}
	p_{\alpha \dot \alpha} = \lambda_\alpha \tilde{\lambda}_{\dot \alpha}
\end{equation}
where $ \lambda_\alpha, \tilde \lambda_{\dot \alpha} $ are two 2-component vectors. The requirement that the gluon momentum be real implies that these two vectors are not independent but satisfy: $ \tilde \lambda_{\dot \alpha} = \pm (\lambda^*_\alpha) $ \cite{Arkani-Hamed2017Scattering}. These $ \lambda^i_\alpha $ are known as the spinor helicity variables.

\subsection{Grassmannian}\label{sec:grassmannian}

Any such set of $ n $ null momenta obeys a certain symmetry which becomes more apparent when we express the initial data in the following manner:
\begin{equation}\label{eqn:n-vectors}
	\begin{pmatrix}
		\bvec{a} \\
		\bvec{b}
	\end{pmatrix} = \begin{pmatrix}
		\lambda^1_1 & \lambda^2_1 & \ldots & \lambda^n_1 \\
		\lambda^1_2 & \lambda^2_2 & \ldots & \lambda^n_2
	\end{pmatrix}
\end{equation}
where we have collected the first (resp. second) component of each of the $ n $ null spinors into an $ n $ dimensional (complex) vector $ \bvec{a} $ (resp. $ \bvec{b} $). Thus the kinematic space for $ n $ massless particle scattering can be specified in terms of two $ n $ dimensional (complex) vectors $ \bvec{a}, \bvec{b} \in \C^n  $. However, this specification is only with respect to a given observer. One can always perform a Lorentz transformation to a different frame of reference where the initial data would be represented by a \textit{different} set of $ n $ dim vectors $ \bvec{a'}, \bvec{b'} \in \C^n $.

There is a simple relation between two such sets of vectors representing the kinematic data in two different Lorentz frames. Since Lorentz transformations act as linear transformations given by elements of $ \sltwoc $, each pair of components of the unprimed and primed vectors will be related to each other by some linear transformation, i.e.:
\begin{align}\label{eqn:linear-sl2}
	a'_i & = \alpha a_i + \beta b_i \nonumber \\
	b'_i & = \gamma a_i + \delta b_i
\end{align}
where the coefficients of the above linear transformations are the elements of the $ \sltwoc $ matrix generating the Lorentz transformation. Consequently the new pair of vectors $ (\bvec{a'}, \bvec{b'}) $ can be written as a linear combination of the old pair of vectors $ (\bvec{a}, \bvec{b}) $. Any two (non-collinear) vectors in $ \C^n $ will span a two-dimensional plane in $ \C^n $. The fact that the Lorentz transformed kinematic data are linear combinations of the original vectors implies that under Lorentz transformations the kinematic data associated with a given $ n $ massless particle scattering amplitude continues to remain in the \textit{same} two-dimensional plane of $ \C^n $. This leads us to the following conclusion:
\begin{quote}
	\textit{The space of kinematic data for the scattering of $ n $ massless particles in Minkowski space consists of the set of all two-planes in $ \C^n $, also known as the Grassmannian $ Gr_{2,n} $.}
\end{quote}

\subsection{Spinor Sums and Momentum Conservation}

Momentum vectors satisfy the following relation: $ 	\sum_{i=1}^n p_i^\mu = 0 $ for each component of momentum $ p^\mu $ individually. This is just the statement of conservation of momentum. Consider, however, that we reverse the time orientation of all the outgoing momenta $ \mb P_{out} $, so that $ p_i^0 \rightarrow - p_i^0 $ where $ p^i_\mu \in \mb P_{out} $. Now, the timelike component becomes:
\begin{equation}\label{eqn:mom-en-sum}
	\sum_{i=1}^{n} p_i^0 = \sum_{i \in \mb P_{in}} p_i^0 - \sum_{i \in \mb P_{out}} p_i^0 = 2 E
\end{equation}
where $ E $ is the total energy involved in the process. In terms of the null spinors defined in \eqref{eqn:null-momenta}, we can write the following expression: $ \ket{\lambda_i}\bra{\lambda_i} = \frac{1}{2} \left( \innerp{\lambda_i}{\lambda_i} \id + \mb { p}_i \cdot \vec \sigma \right) $
where we have used the Dirac notation for the spinors: $ \lambda_i^\alpha \simeq \ket{\lambda_i} = (\lambda_i^0 ~ \lambda_i^1)^T $ and $ \bra{\lambda_i} = (\lambda_i^{0*} ~ \lambda_i^{1*}) = (\ket{\lambda_i}^\dagger) $. Here $ \innerp{\lambda_i}{\lambda_i} = p_i^0 $ is the spinor energy and $ \mb{p}_i $ is the spinor 3-momentum. $ \vec \sigma $ are the usual Pauli matrices. If we were to write the momentum conservation equation in terms of these spinors we would find:
\begin{equation}\label{eqn:total-momentum-v2}
	\ket{\lambda_i}\bra{\lambda_i} = 0.
\end{equation}
The equivalent of \eqref{eqn:mom-en-sum} (with the time orientation of ingoing momenta reversed) would be the expression:
\begin{equation}\label{eqn:spinor-en-sum}
	\sum_{i \in \mb P_{in}} \innerp{\lambda_i}{\lambda_i} - \sum_{i \in \mb P_{out}} \innerp{\lambda_i}{\lambda_i} = 2 E
\end{equation}
for the time-like component, while the spatial components would continue to satisfy:
\begin{equation}\label{eqn:spatial-mom-sum}
	\sum_{i=1}^n \mb p_i = 0
\end{equation}
Thus we can write:
\begin{equation}\label{eqn:total-momentum-v3}
	\sum_{i=1}^{n}\ket{\lambda_i}\bra{\lambda_i} = E \id.
\end{equation}
The reason for reversing the time orientation on the outgoing momenta will become clear shortly.

\section{Coherent States of Quantum Geometry}\label{sec:coherent-states}

\begin{figure}[h]
	\centering
	\begin{subfigure}[t]{0.3\textwidth}
		\centering
		\includegraphics[height=120pt]{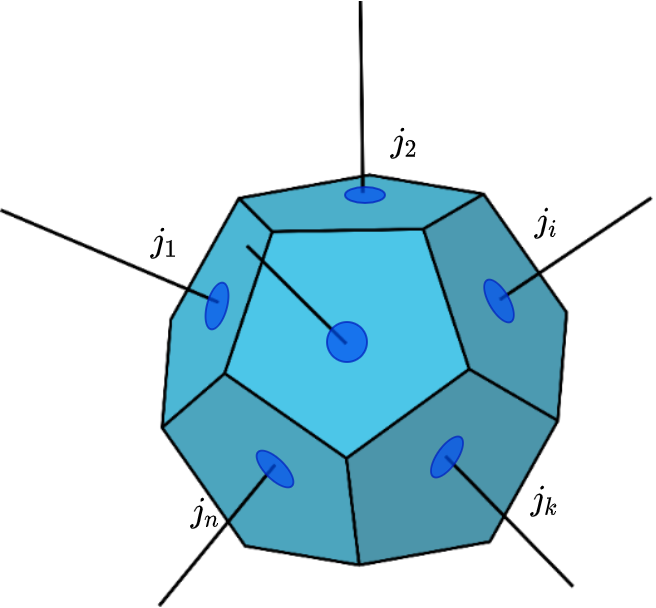}
		\caption{Edges labeled by spins.}
		\label{fig:polyhedron-spin}
	\end{subfigure}
	\hfill
	\begin{subfigure}[t]{0.3\textwidth}
		\centering
		\includegraphics[height=120pt]{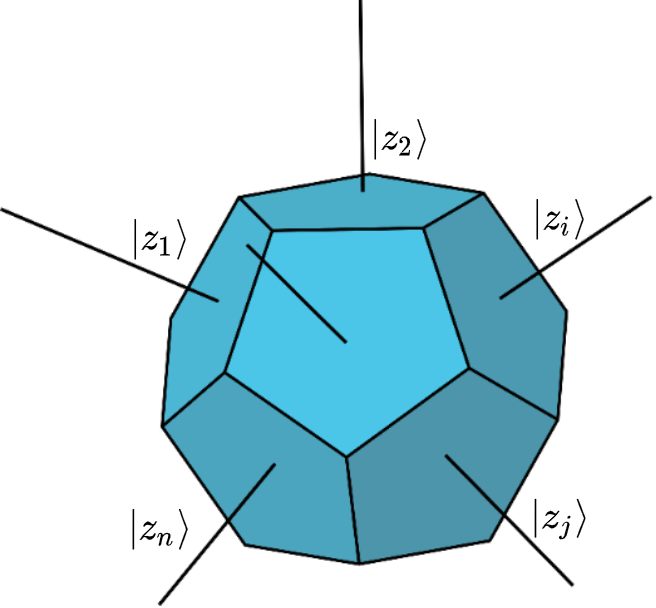}
		\caption{Equivalent description of the vertex state space in of edges labeled by spinors $ z_i \in \mbb C^2 $.}
		\label{fig:polyhedron-spinors}
	\end{subfigure}
	\caption{Single vertex of a spin network represented as a polyhedron.}
	\label{fig:polyhedron}
\end{figure}

\begin{figure}[h]
	\centering
	\includegraphics[width=0.3\linewidth]{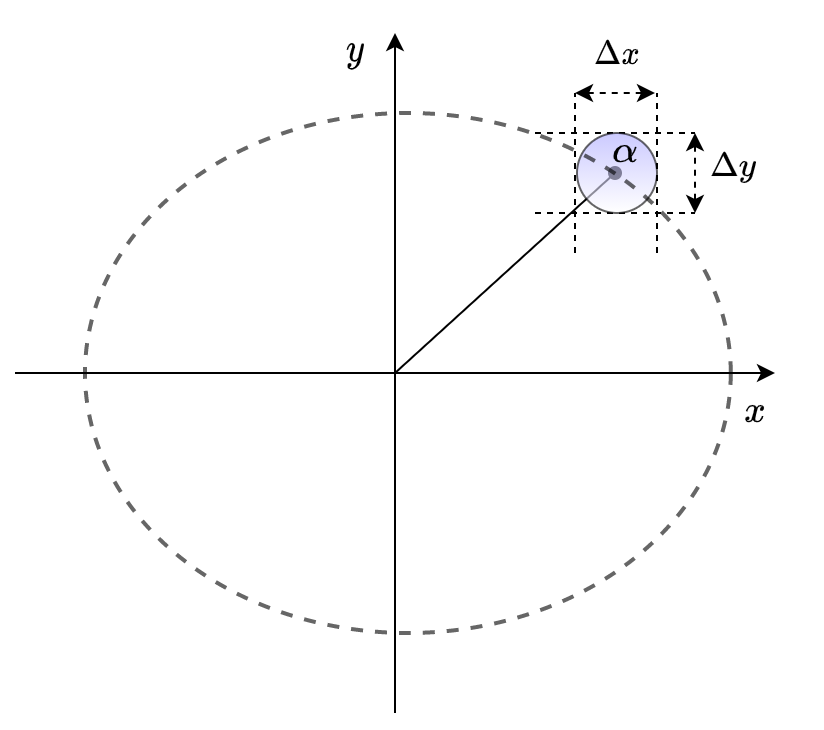}
	\caption{The phase space of a simple harmonic oscillator. The coherent state $ \ket{\alpha} $ saturates the Heisenberg bound $ \Delta x \Delta y = \hslash/2 $. The large ellipse represents the time evolution of $ \ket{\alpha} $ evaluated using Schrodinger's equation and coincides with the trajectory of a classical oscillator. The small circle is the region of uncertainty around the classical phase space point whose co-ordinates are given by $ \expect{\hat x} = \mf{Re}(\alpha),~ \expect{\hat y} = \mf{Im}(\alpha) $.}
	\label{fig:sho-phase-space}
\end{figure}

Loop quantum gravity (LQG) \footnote{See, for instance, \cite{Rovelli2014Covariant} for a detailed introduction and \cite{Vaid2014LQG-for-the-Bewildered} for a more pedagogical presentation suitable for non-experts} is an approach to quantum gravity based on the Hamiltonian formalism of general relativity. Here spacetime is viewed as consisting of a sequence of three-dimensional manifolds, each of which describes the geometry of a given spatial region at a given instant of time. More than 30 years ago, Ashtekar wrote down (see, for instance, \cite{Ashtekar2004Background} for a review) the equations for Hamiltonian general relativity in a form analogous to that used for describing ``gauge'' theories such as electromagnetism and Yang-Mills. It was then possible to apply the wildly successful methods of quantization of quantization of gauge fields based on the concept of a holonomy to the question of quantum gravity. In the resulting framework, states of \emph{quantum} geometry are described by graphs known as spin-networks whose edges are labeled by spins (or more precisely representations of $ SU(2) $) and vertices are labeled by invariant tensors known as \emph{intertwiners} which serve to ``glue'' together all the edges adjacent to that vertex.
Each vertex of a spin-network can be viewed as representing a polyhedron \autoref{fig:polyhedron-spin} whose faces are normal to the edges attached to the given vertex. Spin networks encode the geometry of a three-dimensional manifold in a fairly straightforward manner. Each edge carries a unit of area given by the Casimir $ \sqrt{j(j+1)} $ of the $ SU(2) $ representation living on that edge. Each vertex represents a quantum of volume where the value of the volume depends in a non-trivial way \cite{Rovelli1994Discreteness} on the number of edges coming into that vertex and the set of $ SU(2) $ labels on those edges.
Now, one important test for any theory of quantum gravity is whether one can construct states which resemble smooth spacetime in a suitable limit. In ordinary quantum mechanics one can construct coherent or minimum uncertainty states whose behavior under time-evolution mimics that of classical trajectories as shown for the example of the simple harmonic oscillator in \autoref{fig:sho-phase-space}.

It is natural to ask whether the same is possible in LQG. Can we construct coherent states of quantum geometry which are peaked around classical values of geometrical observables? The answer is in the affirmative. There are several possible constructions of coherent states in LQG, but I will focus on the kind known as $ U(n) $ intertwiner coherent states first developed by Freidel and Livine in \cite{Freidel2009Fine,Freidel2010UN-Coherent}. For this purpose one first has to recast the classical phase of LQG in terms of spinorial variables. In spinorial LQG edges of spin-networks are labeled by spinors $ z_i \in \mbb C^2 $ as shown in \autoref{fig:polyhedron-spinors}. The algebra of geometric operators which act on the spinors associated with a given vertex can be shown to be identical\footnote{The interested reader can refer to the originals papers \cite{Freidel2009Fine,Freidel2010UN-Coherent} for details or to work by the present author \cite{Vaid2021Coherent} for a more compact presentation.} to that of the unitary group $ U(n) $, where $ n $ is the valence (number of attached edges) of the vertex in question. These states take the following form:
\begin{equation}\label{eqn:u-n-coherent-states}
	\bket{J, z_i} = \frac{1}{\sqrt{J+1}}\frac{(F^\dag_\bz)^J}{J!}\bket{0}.
\end{equation}
where $ J $ is the total area of the polyhedron associated with the given vertex, $ \{z_i; i \in 1,\ldots,n\} $ is the set of spinors which label a point in the classical phase space of the vertex, $ F^\dag_\bz $ are a set of creation operators and $ \bket{0} $ is the Fock vacuum of the theory representing the state with zero volume. As in the case of the harmonic oscillator, intertwiner coherent states are labeled by a point in the classical phase space of the system and can be expressed in terms of the action of raising operators on the vacuum of the theory.

The area $ J $ is related to the spinorial variables $ \{z_i\} $ in the following manner:
\begin{equation}\label{eqn:total-area}
	\sum_i \ket{z_i}\bra{z_i} = J \id
\end{equation}
I urge the reader to compare this expression to the one for null momenta \eqref{eqn:total-momentum-v3}, which I recall here for convenience: $ \sum_{i=1}^{n}\ket{\lambda_i}\bra{\lambda_i} = E \id $. From this we can see the total energy involved in the scattering of $ n $ massless particles corresponds to the total area of the polyhedron of a coherent intertwiner with $ n $ faces. This is not all, however. The LQG spinors $ \ket{z_i} $ are null, just like the momentum spinors $ \lambda_i^\alpha $. Moreover the conservation of spatial momentum \eqref{eqn:spatial-mom-sum} corresponds in the LQG language to the statement that 
\begin{equation}\label{eqn:closure-condition}
	\sum_{i=1}^{n} \vec J_i = 0,
\end{equation}
where $ \vec{J}_i $ is the outgoing normal to the $ i^\text{th} $ face of the polyhedron in the classical theory. Geometrically this corresponds to the requirement that the faces of the polyhedron should form a closed surface.

Lorentz transformations act on the LQG spinors in precisely the same way as they do in the case of particle scattering and consequently, for precisely the same reason as in the scattering case, we have the following statement:
\begin{quote}
	\textit{The space of kinematic data for coherent intertwiners in LQG consists of the set of all two-planes in $ \C^n $, or the Grassmannian $ Gr_{2,n} $.}
\end{quote}

\section{Scattering and Quantum Error Correction}

Let us consider a scattering process where $ p $ massless incoming particles scatter to give $ q $ massless outgoing particles, with $ p+q = n $ being the total number of particles involved in the process. According to the discussion in the earlier sections equivalence classes of kinematic data for this process are given by points in the Grassmannian $ Gr_{2,n} $. Different elements in a given equivalence class are related by a Lorentz transformation as described in \eqref{eqn:linear-sl2}.

Now, if we view geometry as an error correcting code then the question arises - what role can error correction play in the description of such a scattering process? In the last section I have shown that there exists a one-to-one correspondence between the kinematic data for massless $ n $ particle scattering and coherent states of quantum geometry. If we take this correspondence seriously then the following picture emerges. Coherent intertwiners of quantum geometry play the role of quantum gates which act on the $ p $ incoming momenta and transform them into the $ q $ outgoing momenta. Each unique configuration of particle scattering (resp. coherent intertwiners) is labeled by a point in the Grassmannian $ Gr_{2,n} $. Each such state is left invariant by the action of $ SL(2,\mb{C}) $ transformations on all the particle momenta (resp. LQG spinors).

In the language of OQECC we have the following identifications. The physical Hilbert space $ \mc{H}_P $ is the configuration space of $ n $ massless particles with arbitrary momenta. Within this space there is a code subspace $ C $ which can further be written as a tensor product space $ C = A \otimes B $. $ A $ is the space of states labeled by points in $ Gr_{2,n} $ and $ B $ represents the Lorentz gauge freedom of each of these states.

Now, any arbitrary state in $ \mc{H}_P $ does not correspond to a valid kinematical configuration. Only the information encoded in the code subspace $ C $ represents the valid kinematical data - those which satisfy momentum conservation \eqref{eqn:spatial-mom-sum} in the scattering picture and the polyhedron closure condition \eqref{eqn:closure-condition} in the intertwiner picture. In this way we have constructed a precise identification between the kinematic data of massless particle scattering with the structure of an operator quantum error correcting code. There are several ways in which this picture is complementary to and also more general than the holographic QECC (HQECC) studied in \cite{Almheiri2014Bulk,Harlow2017The-RyuTakayanagi,Pastawski2015Holographic}. The HQECC framework is non-local and dependent on the AdS/CFT correspondence. The picture given here is local and appears to describe interactions occurring in flat spacetime. Moreover here there is a direct correspondence between matter interactions and geometry, something that is not present in HQECC.

There is, of course, much more work to be done before the correspondence outlined here can be said to be concrete. In particular the following questions remain to be answered:
\begin{enumerate}
	\item What are the error operations $ \mc E $ and the recovery operations $ \mc R $ given in \eqref{eqn:qec-condition} in this picture?
	\item The state space of a quantum polyhedron is left invariant under $ U(N) $ transformations. Coherent intertwiners provide an overcomplete basis for this space. One should be able to provide a precise description of these states in terms of a QECC. What is the precise form of the QECC in terms of the $ U(N) $ algebra?
	\item I have only considered massless particles and ignored the spin and helicity degrees of freedom. Helicity (and therefore the spin) of massless particles is naturally present in the intertwiner picture in terms of a $ U(1) $ redundancy associated with each face of the polyhedron. What happens in the case of massive particles, those with and without spin?
\end{enumerate}
I hope to address these and other questions in future work.

\section*{Acknowledgements}

I have benefited from many fruitful conversations with Devadharshini Suresh. I thank my wife and my family for constant support and encouragement.

\printbibliography

%

\end{document}